\documentclass[12pt]{elsarticle}
\usepackage{amsmath,amssymb}
\usepackage{eepic}
\usepackage{hyperref}
\usepackage{graphicx}
\usepackage{rotating}
\usepackage{caption}
\usepackage{subcaption}
\usepackage{float}
\usepackage{eepic}

\def\al{\alpha}
\biboptions{sort&compress}
\newtheorem{theorem}{Theorem}[section]
\newtheorem{proposition}[section]{Proposition}

\newenvironment{definition}[1][Definition]{\begin{trivlist}
\item[\hskip \labelsep {\bfseries #1}]}{\end{trivlist}}

\usepackage[update,prepend]{epstopdf}
\begin{document}

\begin{frontmatter}

\title{Differences and similarities in the analysis of Lorenz,
Chen, and Lu systems}

\author{G.A. Leonov}
\author{N.V. Kuznetsov}
\ead{nkuznetsov239@gmail.com}

\address{Mathematics and Mechanics Faculty,
St. Petersburg State University, \\ 198504 Peterhof, St. Petersburg, Russia}
\address{Department of Mathematical Information Technology,
University of Jyv\"{a}skyl\"{a}, \\ 40014 Jyv\"{a}skyl\"{a}, Finland \\
nkuznetsov239@gmail.com}

\begin{abstract}
Currently it is being actively discussed the question of
the equivalence of various Lorenz-like systems and
the possibility of universal consideration of their behavior
\cite{Leonov-2013-DAN-ChenLu,Algaba2014758,Algaba2014772,AlgabaEtAll-2013-Chaos,
AlgabaEtAll-2013-PLA,ChenY-2013,Chen-2013}, in view of
the possibility of reduction of such systems to the same
form with the help of various transformations.
In the present paper the differences and similarities in the analysis
of the Lorenz, the Chen and the Lu systems are discussed and it is shown that
the Chen and the Lu systems are valuable for the development of new methods
for the analysis of chaotic systems.
\end{abstract}


\end{frontmatter}

\section{Introduction}

Currently it is being actively discussed the question of
the equivalence of various Lorenz-like systems and
the possibility of universal consideration of their behavior \cite{Leonov-2013-DAN-ChenLu,Algaba2014758,Algaba2014772,AlgabaEtAll-2013-Chaos,AlgabaEtAll-2013-PLA,ChenY-2013,Chen-2013} in view of the possibility of reduction of such systems
to the same form with the help of various transformations.
For example, in the papers \cite{Algaba2014758}:
one can read:
``{\it
  despite the hundreds of works that affirm the contrary,
  we have recently shown that, generically,
  the Chen and the Lu systems are only particular cases of the Lorenz system
}''.

In the present paper the differences and similarities
in the analysis of these systems are discussed and
it is shown that the Chen and the Lu systems are valuable
for the development of new methods for the analysis of chaotic systems.

\section{Lorenz-like systems: Lorenz, Chen, Lu, and Tigan systems}

Consider the famous Lorenz system \cite{Lorenz-1963}
\begin{equation}\label{sys-Lorenz}
 \begin{aligned}
 &\dot x= \sigma(y-x)\\
 &\dot y= \rho x-y+xz\\
 &\dot z=-\beta z+xy,
 \end{aligned}
\end{equation}
where $\sigma,\rho,\beta$ are positive parameters.

Consider the Chen system \cite{ChenU-1999}
\begin{equation}\label{sys-Chen}
 \begin{aligned}
 &\dot x= a(y-x)\\
 &\dot y= (c-a)x+cy-xz\\
 &\dot z=-bz+xy
 \end{aligned}
\end{equation}
and the Lu system \cite{LuChen-2002}
\begin{equation}\label{sys-Lu}
 \begin{aligned}
 &\dot x= a(y-x)\\
 &\dot y=cy-xz\\
 &\dot z=-bz+xy,
 \end{aligned}
\end{equation}
where $a,b,c$ are real parameters.
Systems \eqref{sys-Chen} and \eqref{sys-Lu} are Lorenz-like systems,
which have been intensively studied in recent years.

In 2012 G.A.~Leonov suggested to consider
the following substitutions \cite{Leonov-2013-DAN-ChenLu}
\begin{equation}\label{trans-LEO}
 x \rightarrow hx,\ y \rightarrow hy,\ z \rightarrow hz,\ t \rightarrow h^{-1}t
\end{equation}
with $h=a$. By this transformation for $a\neq 0$ one has in \eqref{sys-Chen} and \eqref{sys-Lu}
\[
 a \rightarrow 1, c \rightarrow \frac{c}{a}, b \rightarrow \frac{b}{a}.
\]
For $a=0$ the Chen and the Lu systems become linear and their dynamics have minor interest.
Thus, without loss of generality, one can assume that $a=1$.
Remark that chaotic parameters,
considered in the works \cite{ChenU-1999,LuChen-2002},
are positive and thus the transformation \eqref{trans-LEO}
with $h=a$ does not change the direction of time.

Later, in 2013, the transformation \eqref{trans-LEO}
was independently considered
in the works \cite{AlgabaEtAll-2013-Chaos,AlgabaEtAll-2013-PLA}\footnote{
submission dates: \cite{Leonov-2013-DAN-ChenLu} --- December 27, 2012;
\cite{AlgabaEtAll-2013-Chaos} ---  22 January 2013;
\cite{AlgabaEtAll-2013-PLA} --- 22 January 2013}  with $h=-c$
for the reduction of the Chen system \eqref{sys-Chen}
\begin{equation}\label{sys-Chen-Lorenz}
 \begin{aligned}
 &
 \begin{aligned}
 &\dot x= -\frac{a}{c}(y-x),\\
 &\dot y= \big(\frac{a}{c}-1\big) x-y+xz,\\
 &\dot z= \frac{b}{c} z+xy,
 \end{aligned}
 &
 \qquad \sigma = -\frac{a}{c}, \quad \rho=\frac{a}{c}-1, \quad
\beta=-\frac{b}{c}
 \quad (\sigma+\rho = -1)
 \end{aligned}
\end{equation}
and the Lu system \eqref{sys-Lu}
\begin{equation}\label{sys-Lu-Lorenz}
 \begin{aligned}
 &
 \begin{aligned}
 &\dot x= -\frac{a}{c}(y-x),\\
 &\dot y= -y-xz,\\
 &\dot z= \frac{b}{c}z+xy,
 \end{aligned}
 &
 \qquad \sigma = -\frac{a}{c}, \quad \rho=0, \quad \beta=-\frac{b}{c}
 \quad (\rho =0)
 \end{aligned}
\end{equation}
to the form of the Lorenz system \eqref{sys-Lorenz}.

Note that here in contrast to the previous transformation:
1) the transformation \eqref{trans-LEO}
with $h=-c$ change the direction of time
for the positive chaotic parameters considered in the works
\cite{ChenU-1999,LuChen-2002},
2) for $c=0$ the Chen and the Lu systems do not become linear and
their dynamics may be of interest.

For $c=0$ in \cite{AlgabaEtAll-2013-Chaos,AlgabaEtAll-2013-PLA}
it is suggested to apply the previous transformation \eqref{trans-LEO} with $h=a$
and is claimed that the Chen and the Lu systems with $c=0$ are
``{\it{a particular case of the T-system}}'' \cite{JiangHB-2010,Tigan-2008} (which was published much later)
\begin{equation}\label{sys-T}
 \begin{aligned}
 &\dot x= a(y-x)\\
 &\dot y=(c-a)x-axz\\
 &\dot z=-bz+xy.
 \end{aligned}
\end{equation}
To fill the formal gap in the notation of the parameters in this case it is required
the additional transformation
$x \rightarrow x/\sqrt{a},\ y \rightarrow y/\sqrt{a},\ z \rightarrow z/a$.
Finally, to transform the Chen and the Lu systems with $c=0$ to the T-system
one has to apply the following transformation
\begin{equation}\label{trans-NK}
 x \rightarrow \sqrt{a}x,\ y \rightarrow \sqrt{a}y,\ z \rightarrow z,\ t \rightarrow a^{-1}t.
\end{equation}

For $\sigma=10, \beta=8/3$ and $0<\rho<1$, the Lorenz system is stable.
For $1<r<24.74\cdots$ the zero fixed point looses its stability and
two additional stable fixed points appear.
For $\rho>24.74\cdots$ all three fixed points become unstable
and trajectories, depending on the initial data,
may be repelled by them in a very complex way.
For the parameter set $\{ \sigma,\beta,\rho\} = \{10,8/3,28\}$
it was found numerically a chaotic strange attractor
in the Lorenz system \cite{Lorenz-1963}.
Various rigorous approaches to the justification of
their existence are based, for example, on
the investigation of instability (hyperbolicity) of trajectories
with the help of computing Lyapunov exponents,
or the computation of fractional Hausdorff dimension.
See also analytical-numerical approach in \cite{Tucker-1999}.

The Chen system with the parameter set $\{a,b,c\} = \{35,3,28\}$
is chaotic \cite{ChenU-1999},
but it may not be chaotic for some other parameter.
The Lu system with the parameter set $\{a,b,c\} = \{36,3,20\}$
is also chaotic \cite{LuChen-2002}
and, likewise, it may not be chaotic for any other parameter.

It is easy to see that a generalized system
\cite{Leonov-2013-DAN-LD,ChenY-2013, Leonov-2013-IJBC}
\begin{equation}\label{sys-Lorenz-gen}
 \begin{aligned}
 &\dot x= \sigma(y-x),\\
 &\dot y= rx-dy-xz,\\
 &\dot z= -bz+xy,
 \end{aligned}
\end{equation}
contains, as special cases, systems \eqref{sys-Lorenz},
\eqref{sys-Chen},
and \eqref{sys-Lu}.
Here again $\sigma > 0$, $b > 0$, but $r$ and $d$ are
certain real parameters.
Note that for the Lorenz system: $d$ = 1,
the Chen system: $d=-c, c>0, r = c-a$,
and the Lu system: $d=-c , c > 0,r = 0$.

As it was noted by one of the reviewers of this paper  {\it ``If we have to give a different name of each planar slice we take in the three-dimensional parameter space of the Lorenz system, we might have a serious problem.''}
On the other hand, we would like to recall the classical 16th Hilbert problem (second part, \cite{Hilbert-1901})
on the number and mutual disposition of limit cycles in two-dimensional polynomial systems,
where one of the tasks is to find the simplest system, from a certain class, 
with the maximum possible number of limit cycles.
We can consider its essential \emph{``chaotic''} analog:
\emph{on the number and mutual disposition of chaotic compact invariant connected sets}
(e.g. local attractors and repellers)
in three- or multi- dimensional polynomial systems.
Many chaotic polynomial systems
have been discovered (e.g., such particular cases of three-dimensional quadratic systems as
the Lorenz, the Rossler, the Sprott, the Chen, the Lu and other systems) and studied over the years,
``{\it but it is not known whether the algebraically simplest chaotic flow has been identified''}
\cite{SprottL-2007,Sprott-2011}.
Thus, one of the attractive feature of Chen and Lu systems is that
the scenarios of transition to chaos in them are similar
to those in the Lorenz system, but, in contrast to the latter,
nonlinear Chen and Lu systems involve only two parameters hence are simpler.

\section{Recent discussion on equivalence of the Lorenz, Chen, and Lu systems}
Recently a very interesting discussion on the equivalence
of the Lorenz, Chen and Lu systems
was initiated in
\cite{AlgabaEtAll-2013-Chaos,AlgabaEtAll-2013-PLA,Chen-2013}.

Below a few remarks, concerning the discussion
and being important, are given.

1) The Lorenz system is the system considered in the
original work by Edward Lorenz \cite{Lorenz-1963}:
{\it Lorenz, E. N. (1963). Deterministic nonperiodic flow.
J. Atmos. Sci., 20(2):130-141}.
E. Lorenz obtained his system as a truncated model
of thermal convection in a fluid layer
and the parameters $\sigma$, $\rho$, and $b$
of his system are positive
because of their physical meaning 
(e.g., $b = 4(1+a^2)^{-1}$ is positive and bounded).
Thus, from a physical point of view, systems
\eqref{sys-Chen-Lorenz} and \eqref{sys-Lu-Lorenz}
are not particular cases of the Lorenz system
since $\beta$ is negative in \eqref{sys-Chen-Lorenz} and \eqref{sys-Lu-Lorenz}
for the positive $b$ and $c$.

Formally to try to preserve the physical meaning of parameters
one may compare systems \eqref{sys-Chen-Lorenz} and \eqref{sys-Lu-Lorenz} 
for non-positive parameters\footnote{
\eqref{sys-Chen-Lorenz} for $\frac{a}{c}-1<0$;
$\rho=0$ for both forward and backward time in \eqref{sys-Lu-Lorenz}.
} with the Lorenz system in
backward time (time-reversal Lorenz system).
But backward time does not have 
a clear physical sense for the Lorenz system as well
as for many other physical problems.
If one would consider backward physical time,
then it would be logical to name the Lorenz system,
in the backward chronological order,
as the generalization of \emph{time-reversal Chen system}
or \emph{time-reversal Lu system}.

2) From a mathematical point of view,
one may consider nonpositive parameters or backward time.
As it was noted by one of the reviewers of this paper
``{\it to know its full dynamical behavior it is enough
(in the case $c>0$)
to reverse the time in the corresponding
dynamical behavior of the Lorenz system ...''}.

In fact, a consideration of system
in the backward time seems to be needless
if all the objects of interest and their properties can be
obtained from the study of this system in the forward time.
As rightly noted in the works
\cite{AlgabaEtAll-2013-Chaos,AlgabaEtAll-2013-PLA},
the existence of periodic or homoclinic trajectories 
can be studied in one of the time directions.
However for the study of non-closed trajectories
and the sets of such trajectories
being invariant in the forward time,
it may be not the case.

For example, the definition of a dynamical system and
the consideration of limit behavior of trajectories require a proof of
trajectory existence.
Generally speaking, for quadratic systems
the existence of a trajectory on
$t \in [t_0,+\infty)$ does not imply its existence on $t \in (-\infty,t_0]$
(e.g., consider the classical example $\dot x = x^2$ or
multidimensional examples from the paper \cite{GingoldS-2011}
on the completeness of quadratic polynomial systems).

Note that in \cite{AlgabaEtAll-2013-Chaos,AlgabaEtAll-2013-PLA}
there is no discussion of the following important questions for 
the consideration of the Lorenz system in the backward time
or with negative parameters:
the existence of the extension of solutions,
the existence of attractors,
and the possibility of consideration of invariant sets in the backward time.
Some necessary results can be found in
\cite{Coomes-1989}, \cite[p. 35]{Chueshov-2002-book}. 
See also \cite{QinC-2007}.
However they differ from similar consideration
for the Lorenz system in the forward time.

In \cite{AlgabaEtAll-2013-Chaos} one can read:
``{\it{Chen's attractor exists if Lorenz repulsor exist}}'' and
``{\it{most of the literature on the Chen system
is redundant because the results obtained can be
directly derived from the corresponding results on the Lorenz system}}''.
But the question of importance is what was known about
repulsor (or repeller) in the Lorenz system and
the dynamics of time-reversal Lorenz system
before the works \cite{ChenU-1999,LuChen-2002} were published?
To the best of our knowledge even visualization and localization of
the Lorenz repulsor had been unknown.
Since corresponding results are not discussed in
\cite{AlgabaEtAll-2013-PLA,AlgabaEtAll-2013-Chaos}
it would be appropriate to add that
\emph{some of the literature on the Chen and the Lu systems are new and of interest
because the results obtained cannot be directly derived from
the corresponding results on the Lorenz system},
since corresponding results on the Lorenz system have been unknown.

Recall that in the case $c=0$
in \cite{AlgabaEtAll-2013-Chaos,AlgabaEtAll-2013-PLA,Chen-2013}
there is remarked that
the Chen and the Lu systems are ``{\it only a particular case of the T-system}'',
which was published in 2004 \cite{Tigan-2008}
(i.e. later than the Chen and the Lu systems were published).

Remark also that even in the case of the existence of the corresponding objects
in forward and backward time their characteristics can be substantially different.

For example, in general, absolute values of
Lyapunov exponents of a bounded trajectory
in forward and backward time can be quite different.
Also, widely used Kaplan-Yorke dimension (or Lyapunov dimension) of an invariant set
can be defined only for one direction of time.

3) Besides the fact of simultaneous existence
of the corresponding objects in forward and backward time
(being equivalent: closed orbits, homoclinic orbits,
invariant sets and others; or dual: attractor --- repeller),
it is of importance the possibility to find the object
and to analyse its properties in forward or backward time.
In this case the questions of importance arise:
a) whether
the methods, developed for the study of a system in forward time,
can be applied in a similar way to the study of a system in backward time;
and
b) whether a universal consideration of a system
in both forward and backward time is possible.

Next we consider some differences and similarities
in the analysis of the above mentioned systems
and demonstrate that for the study of some properties of
time-reversal Lorenz,  Chen, and Lu systems new methods are needed.

\section{Differences and similarities in the analysis
of the considered systems}

\subsection{Homoclinic orbits} 

 Consider the Lorenz system with fixed parameters $\sigma=10$ and $\beta= 8/3$
 and a varying parameter $\rho$,
following the works of E.~Lorenz.
 For $0<\rho<1$ the origin is a globally stable fixed point.
 Then, for $\rho>1$ the origin becomes unstable
 and two new stable fixed points arise,
 the basins of attraction of which are separated by the stable manifold
of the unstable origin.
 For $\rho=13.9...$ this stable manifold contains a homoclinic orbit
\cite{KaplanY-1979-CMP}.
This result was generalized by G.A.~Leonov.

 \begin{theorem} \cite{Leonov-2009-CAMQ,Leonov-2012-PLA,Leonov-2013-AMM}
 \label{Theorem2}
 Let the numbers $\beta$ and $\sigma$ be given. For the existence
 of $\rho>1$ such that system \eqref{sys-Lorenz} has a homoclinic
trajectory it is necessary and sufficiently that
 \begin{equation}\label{eqno7}
 2\beta+1<3\sigma.
 \end{equation}
 \end{theorem}

 The sufficiency of condition \eqref{eqno7}
 was first obtained in \cite{Leonov-1988,Leonov-1988-RMS}.
 The hypothesis that inequality \eqref{eqno7} is a necessary
condition was accepted
 in \cite{Leonov-1988,Leonov-1988-RMS} and was first proved in
\cite{Chen-1996}.

\begin{figure}[h]
 \includegraphics[width=0.32\textwidth]{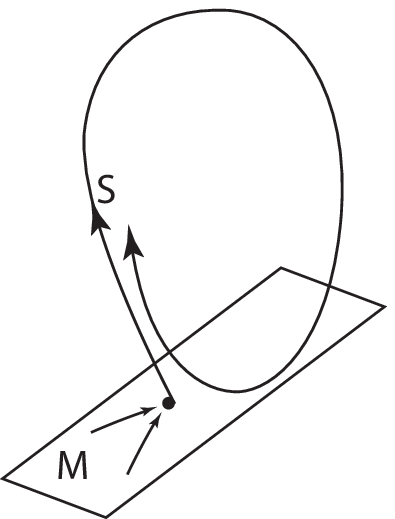}
 \hfill
 \includegraphics[width=0.32\textwidth]{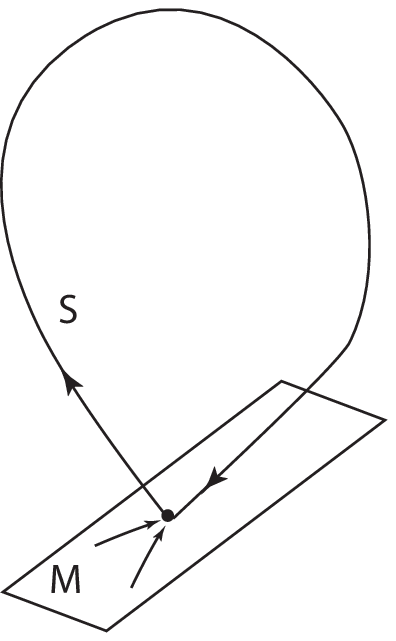}
 \hfill
 \includegraphics[width=0.32\textwidth]{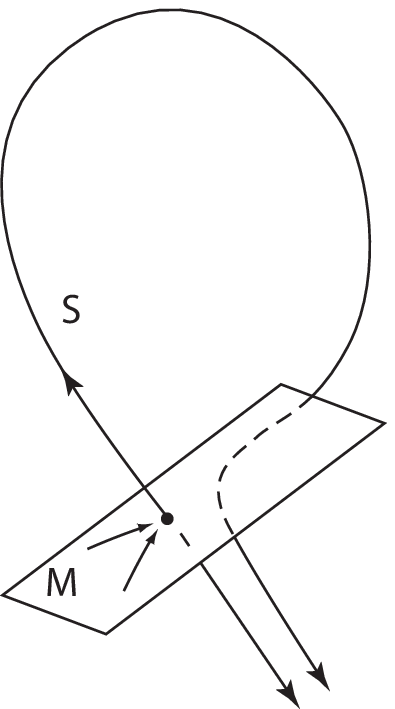}
 \caption{Bifurcation of the birth of a homoclinic orbit}
 \label{homoclinic-bif}
\end{figure}

 Recently in the papers \cite{Leonov-2012-PLA,Leonov-2013-IJBC}
 there is proposed
 a new effective analytical-numerical procedure
 for localization of homoclinic trajectories (Fishing principle).
 For applying this method to three-dimensional systems
 it is of very importance the existence of the two-dimensional stable manifold
 of a saddle point,
 on which the trajectories are attracted to the saddle point
 from which a homoclinic trajectory is outgoing (see Fig.~\ref{homoclinic-bif}).
 For the computation of a homoclinic trajectory
 the initial data in numerical integration
 are chosen closely to a saddle point and its unstable one-dimensional manifold,
 for example, on the eigenvector
 corresponding to the positive eigenvalue of the saddle.
 The purpose of numerical procedures is to reveal
 when the outgoing trajectory returns to the stable two-dimensional manifold
 (see Fig.~\ref{homoclinic-bif}).

 Using Fishing principle, one can obtain \cite{Leonov-2013-DAN-ChenLu,Leonov-2012-PLA,Leonov-2013-IJBC}
 the following approximations
for the Lorenz system with the parameters $\sigma=10, \beta=8/3$, and
$$
\rho \in [13.92, 13,93],
$$
for the Lu system with the parameters $r=0, \sigma=35, d=-28$, and
$$
 b\in [44.963, 44.974],
$$
and for the Chen system with the parameters $r=-7, d=35, d=-28$, and
$$
 b\in[40.914, 40.935].
$$

Though the inversion of time does not affect on
the existence of the homoclinic trajectory,
it makes impossible effective application of modern analytical
and numerical methods to the proof of the existence of homoclinic trajectories.

 \begin{figure}[h]
 \centering
 \includegraphics[width=0.32\textwidth]{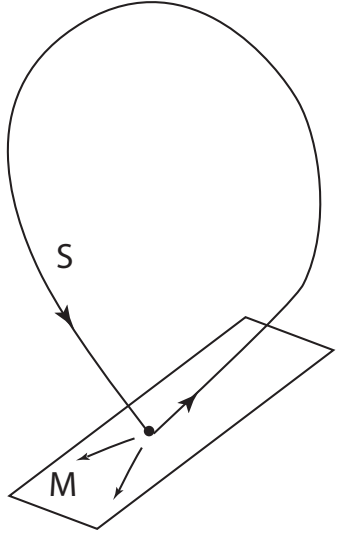}
 \caption{Invertation of time and homoclinic orbit}
 \label{invert}
\end{figure}

Here (see Fig.~\ref{invert})
because of the existence of the two-dimensional unstable manifold of a saddle:
1) in numerical procedure for each parameters set
it is necessary to consider a set of initial data
in a neighborhood of the saddle point
close to two-dimensional unstable manifold
(for example, on a plane, spanned on
 two eigenvectors corresponding to positive
 eigenvalues of the saddle),
2) while only one trajectory, corresponding to the homoclinic
orbit, returns back to the considered unstable manifold,
while the rest of trajectories are repelled by this manifold.
Therefore in the numerical analysis of trajectories
with the above-mentioned initial data
it may be considered a substantial nonuniformity in their behavior,
what makes more difficult a numerical analysis
(for example, for the Lorenz system in a neighborhood
of zero saddle point, for $x=y=0$ the trajectories of system
are exponentially repelled by this point: $z(t) \approx e^{\beta t}$).
The above shows the significant multiple increasing of computational complexity
and the impossibility of effective numerical analysis.

 The existence, in a system, of homoclinic trajectories is
 an valuable tool in studying chaos.
 So-called Shilnikov chaos exists in three-dimensional dynamical
 systems with a homoclinic trajectory of
 a saddle point of equilibrium
if this
 equilibrium is a saddle--focus with a positive saddle value
\cite{ShilnikovTCh-2001}.
 While for the Lorenz, the Chen, and the Lu systems
 there are parameters, corresponding to a homoclinic trajectory
 of zero equilibrium, but
 this zero state is not a saddle--focus.
 Nevertheless in \cite{Leonov-2013-IJBC} it is shown that a small change of
 all these systems in a neighborhood of a saddle can lead to the
satisfaction of all conditions of the Shilnikov theorem and, consequently,
to Shilnikov chaos in the Lorenz, the Chen, and the Lu systems.
Such a construction requires also to make the use of an analog
of Fishing principle and the existence of a two-dimensional stable manifold.

\subsection{Divergence and attractor dimension}

Consider a dynamical system
\begin{equation}\label{dn-sys}
 \frac{dx}{dt}=f(x),\quad x\in \mathbb{R}^n,\, f \in \mathbb{C}^1
\end{equation}
and its linearization along the solution $x(t)=x(t,x_0)$ for $t
\in [0,+\infty)$,
\begin{equation}\label{dl-sys}
 \frac{dy}{dt}=J(t)y, \quad
 J(t)=J(x(t,x_0))=
 \bigg\{
 \frac{\partial f_i(x)}{\partial x_j}\bigg|_{x=x(t,x_0)}
 \bigg\}
\end{equation}
Denote by $\lambda_1(x(t,x_0))\ge\ldots\ge\lambda_n(x(t,x_0))$ the eigenvalues
of the symmetric Jacobi matrix $(J(x(t,x_0)) + J(x(t,x_0))^*)$.

Important property of dynamical system \eqref{dn-sys}
its divergence
\begin{equation}\label{div-def}
 {\rm div} f(x(t,x_0))= \sum\limits_{1}^{n}
 \frac{\partial f_i(x)}{\partial x_i}|_{x=x(t,x_0)}=
 \dfrac{1}{2}\sum\limits_{1}^{n} \lambda_i(x(t,x_0)),
\end{equation}
which characterizes the change of volume
along the trajectory $x(t,x_0)$
in the phase space.

Suppose $X(t)$ is a fundamental matrix of system \eqref{dn-sys}
and $\al_1(X(t))\ge \cdots \ge \al_n(X(t)) \ge 0$ are its singular values
(the square roots of eigenvalues of the matrix $X(t)^*X(t)$ are renumbered for each $t$).
Geometrically, the values $\al_j(X(t))$ coincide with the
principal axes of the ellipsoid $X(t)B$, where $B$ is a unit
ball.
The Lyapunov exponent $\mu_j$ at the point $x_0$ is a number (or the symbol -$\infty$
or +$\infty$):
$$
 \mu_j(x_0)= \lim\limits_{t \to +\infty}\sup\frac{1}{t}\ln\al_j(X(t)).
$$
By definition, $\mu_j(x_0)$ is the \emph{exact Lyapunov exponent} if there exists
a finite limit
$
 \lim\limits_{t \to +\infty}\frac{1}{t}\ln\al_j(X(t)).
$

If the Lyapunov exponents of system exist and are finite, then
\begin{equation}\label{div-def}
 \lim\limits_{t \to +\infty}\sup
 \frac{1}{t} \int\limits_{0}^{t}{\rm div} f(x(\tau,x_0))d\tau
 = \sum\limits_{1}^{n} \mu_j(x_0).
\end{equation}

\begin{proposition} \cite{LeonovPS-1996}
 If ${\rm div} f(x(t,x_0)) > 0$,
 then a stationary solution $x(t,x_0) \equiv x_0$ is Lyapunov unstable
 and a periodic trajectory $x(t,x_0)=x(t+T,x_0)$ is orbitally unstable\footnote{
 For an arbitrary solution it, generally speaking, is not valid
 (see, e.g., a gap in the proof of the Chetaev theorem on instability
 by the first approximation \cite{LeonovK-2007, Leonov-2008}).
}.
\end{proposition}

For the Chen systems \eqref{sys-Chen} and \eqref{sys-Chen-Lorenz}
and for the Lu systems \eqref{sys-Lu} and \eqref{sys-Lu-Lorenz}
the divergence is constant
and under the condition $a+b>c$ one has
\begin{equation}\label{sys-Chen-div}
 {\rm div} = -a+c-b< 0
\end{equation}
and
\begin{equation}\label{sys-Chen-Lorenz-div}
 {\rm div} = \frac{a}{c}-1+\frac{b}{c} > 0,
\end{equation}
respectively.
Therefore systems
\eqref{sys-Chen-Lorenz} and \eqref{sys-Lu-Lorenz},
unlike \eqref{sys-Chen} and \eqref{sys-Lu},
are not dissipative in the sense ${\rm div} < 0$
and there occurs only volumes increasing.
Consequently the bounded set cannot be positively invariant and
it is not dissipative in the sense of Levinson \cite{LeonovBSh-1996}.

The idea of volume contracting
is a base of the dimension theory of attractors\footnote{
Following \cite{Broer-1991,Leonov-2008}, it can be found that an attractor
is a bounded, closed, invariant attracting subset
of the phase space of a dynamical system.
The different types of attraction and
rigorous definitions of attractors can be found
in \cite{BoichenkoLR-2005,Milnor-2006}.
}
(see, e.g., \cite{DouadyO-1980,Temam-1988,BoichenkoLR-2005,Leonov-2012-PMM}).
For finite-dimensional dynamical systems
it was possible to try to get nontrivial results:
in this case the estimates of dimension might not be less
than the dimension of the phase space.
Then it was essential to try to extend the well-known Liouville theorem
on the volume $V(K)$ contracting for the invariant compact set $K\subset U$
of differential equation.

\begin{theorem}
If
\begin{equation}\label{eq-2}
 {\rm div}\,f(x)<0,\quad \forall\,x\in U \subset \mathbb{R}^n,
\end{equation}
then $V(K)=0$.
\end{theorem}

The mathematical tools, developed independently by A.~Douady \& J.~Oesterle \cite{DouadyO-1980}
and Yu.S.~Ilyashenko \cite{Ilyashenko-1982},
permitted one to obtain the following extension of the Liouville theorem
and to estimate the Hausdorff dimension of $K$. 

\begin{theorem}
Suppose that the inequality
\begin{equation}\label{eq-3}
 \lambda_1(x)+\cdots+\lambda_k(x)+s\lambda_{k+1}(x)<0,\,\forall x\in
U, s\in[0,1]
\end{equation}
is satisfied.
Then the Hausdorff dimension of an invariant compact set $K\subset U$
has the following estimate
\begin{equation}\label{eq-4}
 \dim_HK<k+s
\end{equation}
\end{theorem}
Remark that condition \eqref{eq-3} can be satisfied
only if the divergence is negative.

In the work \cite{Leonov-1991-Vest}
it is introduced Lyapunov functions in the estimates of the form \eqref{eq-3}
and proved the following result.

\begin{theorem} \cite{Leonov-1991-Vest,Leonov-2012-PMM}
If there exists a differentiable function $v(x)$ such that
the inequality
$$
\lambda_1(x)+\cdots+\lambda_k(x)+s\lambda_{k+1}(x)+\dot
v(x)<0,\,\forall x\in U
$$
is satisfied, then estimate \eqref{eq-4} is valid.
\end{theorem}

Nowadays the various characteristics of attractors of dynamical systems
(information dimension, metric entropy, etc)
are studied based on Lyapunov exponents computation\footnote{
While positive largest Lyapunov exponent, computed along a trajectory,
is widely used as indication of chaos,
the rigorous mathematical consideration
requires the verification of additional properties
(e.g., such as regularity, ergodicity)
since there are known the Perron effects of the largest Lyapunov exponents sign-reversal
for non-regular linearization \cite{KuznetsovL-2001,LeonovK-2007,KuznetsovL-2005}.
The regularity of almost all linearizations of a dynamical system
and the existence of exact limits of Lyapunov exponents (for almost all $x_0$)
with respect to an invariant measure
result from the Oseledets theorem \cite{Oseledec-1968}.
However in the general case there are no effective methods
for the construction of an invariant measure in a phase space of a system,
the support of which is sufficiently dense.
 More essential justification of the existence of exact values of Lyapunov exponents
 in computer experiments may be the following:
 in calculations with finite precision
 any bounded pseudo-trajectory $\widetilde{x}(t,x_0)$ has
 a point of self-intersection:
 $\exists t_1, t_2: \widetilde{x}(t_1,x_0) = \widetilde{x}(t_1+t_2,x_0))$.
 Then for sufficiently large $t\geq t_1$ the trajectory
 $\widetilde{x}(t,x_0)$ may be regarded as periodic.
 From a theoretical point of view this fact is relies on the shadowing theory,
 the closing lemma, and its various generalizations
 (see, e.g., the surveys \cite{Mane-1984,Pilyugin-2011,Hertz-2013,Sambarino-2014}).
}.

\emph{Local Lyapunov dimension} of a point $x_0$
is defined by
\begin{gather}\label{formula:kaplan}
 \dim_Lx_0 = j(x_0) + \cfrac{\mu_1(x_0) + \ldots +
\mu_j(x_0)}{|\mu_{j+1}(x_0)|},
\end{gather}
where
$\mu_1(x_0)~\geq~\ldots~\geq~\mu_n(x_0)$
are Lyapunov exponents;
$j(x_0) \in [1, n]$ is the smallest natural number $m$ such that
$$
\mu_1(x_0) + \ldots + \mu_{m+1}(x_0) < 0, \quad
\mu_{m+1}(x_0) < 0, \quad
\cfrac{\mu_1(x_0) + \ldots + \mu_m(x_0)}{|\mu_{m+1}(x_0)|}
< 1.
$$

The Lyapunov dimension of an invariant set $K \subset U$ of a dynamical system
is defined as
\begin{equation}
 \dim_LK = \sup_{x_0 \in K} \dim_Lx_0.
\end{equation}

\emph{Lyapunov dimension} is an estimate from above
of topological, Hausdorff, and fractal dimensions \cite{Hunt-1996,BoichenkoLFR-1998,BoichenkoLR-2005}
$$
 \dim_TK\le\dim_HK\le\dim_FK\le\dim_LK.
$$

The estimate from above of the Lyapunov dimension by Lyapunov functions
\cite{Leonov-2013-DAN-LD}
and its comparison with the local Lyapunov dimension in zero stationary point
permit one to obtain the exact formula of dimension
for a generalized Lorenz system \eqref{sys-Lorenz-gen} with a certain parameter $d$.
For example, for the Lorenz system (where $d=1$)
the following result is known:

\begin{theorem} \cite{LeonovPS-2011,LeonovPS-2013-PLA,Leonov-2013-DAN-LD}
If all the equilibria of the Lorenz system are hyperbolic, then
\[
 \dim_LK= 3-\frac{2(\sigma+b+1)}{\sigma+1+\sqrt{(\sigma-
1)^2+4r\sigma}}.
\]
\end{theorem}

However the case $d<0$ is turned to be more complicated
and a similar consideration does not permit one to obtain
similar expressions of the attractors of the Chen and the Lu systems
for classical parameters (where $d=-1$).
Here the assertion on a coincidence of the Lyapunov dimension of attractors
with its value at zero stationary point is a conjecture,
the proof of which requires the construction of new Lyapunov functions
and a careful consideration\footnote{
 For example, in \cite{ChenY-2013} there are given
 the results on the analysis of the generalized Lorenz system \eqref{sys-Lorenz-gen}.
 In the case $d>0$ these results are special cases of more general
 results, published earlier in \cite{Leonov-2013-DAN-LD}.
 In the case $d<0$ the authors' students N.~Korzhemanova and D.~Kusakin revealed a gap
 in the reasoning given in \cite{ChenY-2013}.
 They found parameters, under which the validity of condition (55)
 does not imply the validity of condition (49):
 $a=1, b=2, c=6, d=1$, and $\gamma_1 = 5, \gamma_2 =-59/12$.
}\footnote{
 In \cite{KuznetsovMV-2014-CNSNS} Leonov's conjecture on the Lyapunov
dimension of the Rossler attractor was verified numerically,
while an analytical proof is still an open problem.
}.

Note also that the time inversion (e.g. in the change
\eqref{trans-LEO} with time reversal)
may lead to quite different values of Lyapunov values
and positive divergence
(in this case it is impossible to introduce a nontrivial Lyapunov dimension and to estimate it).
In general,
\[
 \limsup_{\tau \to +\infty}\frac{1}{\tau}\ln|x(\tau)| = -\liminf_{t \to -\infty}\frac{1}{t}\ln|x(-t)|.
\]

\subsection{Application of dimension estimates to the problem on stability of
stationary sets. Analog of Bendixson criterion}

Consider a certain set $D\subset \mathbb{R}^n$, diffeomorfic to
closed ball, the boundary of which $\partial D$ is transverse
to the vectors $f(x),\,\,x\in\partial D$. In this case $D$
is positively invariant for the solutions $x(t)$ of system
\eqref{dn-sys}.

\begin{theorem}\label{thm11}\cite{Leonov-1991,LeonovB-1992}.
Suppose that there exists a continuously differentiable function $v(x)$
and a nondegenerate matrix $S$ such that
\begin{equation}\label{eq-36}
 \lambda_1(x,S)+\lambda_2(x,S)+\dot v(x)<0,\,\,\forall\,x\in D.
\end{equation}
Then any solution $x(t)$ of system \eqref{dn-sys} with the initial data
$x(0)\in D$ tends for $t\to+\infty$ to a stationary set.
\end{theorem}

From Theorem \ref{thm11} it follows at once the following.

\begin{theorem}\label{thm12}
Suppose that there exists a continuously differentiable function $v(x)$
such that
$$
 \lambda_1(x,S)+\lambda_2(x,S)+\dot v(x)<0,\,\,\forall\,x\in
\mathbb{R}^n.
$$
Then any bounded for $t\ge 0$ solution of system (9) tends
for $t\to+\infty$ to a stationary set.
\end{theorem}

For the Lorenz system \eqref{sys-Lorenz}, condition \eqref{eq-36} is satisfied for
\begin{equation}\label{eq-39}
 r<(b+1)(\frac{b}{\sigma}+1).
\end{equation}

\subsection{Numerical simulation and visualization of attractors}

An oscillation in a dynamical system can be easily localized numerically
if the initial conditions from its open neighborhood lead to long-time behavior
that approaches the oscillation.
Thus, from a computational point of view it is natural to suggest the following classification of attractors,
based on the simplicity of finding the basin of attraction in the phase space:

\begin{definition}\cite{KuznetsovLV-2010-IFAC,LeonovKV-2011-PLA,LeonovKV-2012-PhysD,LeonovK-2013-IJBC}
 An attractor is called a \emph{hidden attractor} if its
 basin of attraction does not intersect with
 small neighborhoods of equilibria,
 otherwise it is called a \emph{self-excited attractor}.
\end{definition}

For a \emph{self-excited attractor} its basin of attraction
is connected with an unstable equilibrium
and, therefore, self-excited attractors
can be localized numerically by the
\emph{standard computational procedure},
in which after a transient process a trajectory,
started from a point of an unstable manifold in a neighborhood
of an unstable equilibrium,
is attracted to the state of oscillation and traces it.
Thus self-excited attractors can be easily visualized.

In contrast, for a hidden attractor its basin of attraction
is not connected with unstable equilibria.
For example, hidden attractors are attractors in
the systems with no equilibria
or with only one stable equilibrium
(a special case of multistable systems and
coexistence of attractors)\footnote{
e.g., the nested limit cycles in the papers on the 16th Hilbert problem,
counterexamples to the Aizerman and Kalman conjectures
on the absolute stability of nonlinear control systems,
and others examples.
}.
Recent examples of hidden attractors can be found in
\citep{LeonovKKSZ-2014,ZhusubaliyevM-2014,LiSprott-2014-IJBC,WeiML-2014-TJM-cu,ZhaoLD-2014,
LaoSJS-2014-cu,ChaudhuriP-2014-cu,LiZY-2014-cu,PhamJVWG-2014,PhamRFF-2014}.
Multistability is often an undesired situation in many applications,
however coexisting self-excited attractors
can be found by the standard computational procedure.
In contrast, there is no regular way to predict
the existence or coexistence of hidden attractors in a system.
Note that one cannot guarantee the localization of an attractor
by the integration of trajectories with random initial conditions
(especially for multidimensional systems)
since its basin of attraction may be very small.

 Classical Lorenz, Chen, and Lu attractors are self-excited
 attractors, and consequently they can be easily found numerically.
 If E.~Lorenz, a pioneer of chaos theory,
 studied his system with inverted time by a reason of instability,
 he would not find by numerical experiments his famous attractor,
 which became repeller in the case of inverted time,
 and the theory of chaos would come into being much later.
%
%

\section{Conclusion}

 In the present paper we considered the difficulties of investigation
of Lorenz-like systems, related to inversion of time or negativeness of parameters.
For example, the changes of variables with time inversion,
reducing the Chen and the Lu systems for a certain set of parameters
to the form of the Lorenz system, make impossible
the application of effective numerical procedures for attractor vizualization
(since an attractor becomes a repeller) and the analysis of its dimension,
the development of effective analytical methods
for the study of chaotic behavior and characteristics of attractors:

 1) It makes impossible the effective application of modern analytical and numerical methods
 to the proof of the existence of homoclinic trajectories

 2) The transformed Chen \eqref{sys-Chen-Lorenz} and Lu \eqref{sys-Lu-Lorenz} systems (with time inversion),
 unlike the original Chen \eqref{sys-Chen} and Lu \eqref{sys-Lu},
 are not dissipative in the sense ${\rm div} < 0$
 and there occurs only volumes increasing.
 Consequently a bounded set cannot be positively invariant and they
 are not dissipative in the sense of Levinson.

3) Time inversion, used for the reduction of Chen and Lu systems
 to the Lorenz system, makes it impossible to introduce a Lyapunov dimension
 and to estimate it.

\section*{\uppercase{Acknowledgements}}
 The authors thank Alejandro~J.~Rodriguez-Luis,
 Julien Clinton Sprott for the interesting discussion.
 This work was supported by Russian Scientific Foundation (project 14-21-00041)
 and Saint-Petersburg State University.


\end{document}